# A graph neural network-based multispectral-view learning model for diabetic macular ischemia detection from color fundus photographs

Qinghua He[1,2,‡], Hongyang Jiang[1,‡,*], Danqi Fang[1], Dawei Yang[1], Truong X. Nguyen[1], Anran Ran[1,2], Clement C. Tham[1,2], Simon K. H. Szeto[1,3], Sobha Sivaprasad[4], Carol Y. Cheung[1,2,*]

[1]Department of Ophthalmology and Visual Sciences, The Chinese University of Hong Kong, Hong Kong Special Administrative Region, China
[2]Lam Kin Chung. Jet King-Shing Ho Glaucoma Treatment and Research Centre, Department of Ophthalmology and Visual Sciences, The Chinese University of Hong Kong, Hong Kong Special Administrative Region, China
[3]Hong Kong Eye Hospital, Hong Kong Special Administrative Region, China
[4]NIHR Moorfields Biomedical Research Centre, Medical Retina, Moorfields Eye Hospital, London, EC1V 2PD, UK
[‡]These authors contributed equally to this work and should be regarded as co-first authors.
*Email: hongyangjiang@cuhk.edu.hk, carolcheung@cuhk.edu.hk.

## Abstract

Diabetic macular ischemia (DMI), marked by the loss of retinal capillaries in the macular area, contributes to vision impairment in patients with diabetes. Although color fundus photographs (CFPs), combined with artificial intelligence (AI), have been extensively applied in detecting various eye diseases, including diabetic retinopathy (DR), their applications in the detection of DMI remain unexplored, partly due to the skepticism among ophthalmologists regarding its feasibility. In this study, we propose a graph neural network-based multispectral-view learning (GNN-MSVL) model designed to detect DMI from CFPs. The model leverages higher spectral resolution to capture subtle changes in fundus reflectance caused by the ischemic tissue, thereby enhancing sensitivity to DMI-related features. The proposed approach begins with computational multispectral imaging (CMI) to reconstruct 24-wavelength multispectral fundus images from CFPs. ResNeXt101 is then employed as the backbone for multi-view learning to extract features from the reconstructed images. Additionally, a GNN with a customized jumper connection strategy is designed to enhance cross-spectral relationships, facilitating comprehensive and efficient multispectral-view learning. The study included a total of 1,078 macula-centered CFPs from 1,078 eyes of 592 patients with diabetes, of which 530 CFPs from 530 eyes of 300 patients were diagnosed with DMI. The model achieved an accuracy of 84.7% and an area under the receiver operating characteristic curve (AUROC) of 0.900 [95% CI: 0.852–0.937] on eye-level, outperforming both the baseline model trained from CFPs and human experts (p-values <0.01). These findings suggest that AI-based CFPs analysis holds promise for detecting DMI, which may contribute to its early and low-cost screening.

**Keywords:** Color fundus photograph, diabetic macular ischemia, computational multispectral imaging, graph neural network, multi-view learning

## Introduction

DMI is a condition marked by a loss of retinal capillaries in the macular area and the enlargement and irregularity of the foveal avascular zone (FAZ) in patients with diabetes mellitus (DM)[1,2]. DMI is recognized as a risk factor for the progression of DR and worsening vision, so it has been gradually considered as a possible new therapeutic target to halt DR progression and visual deterioration[3-6]. Multiple factors in DM contribute to the impairment of retinal capillaries, manifesting as capillary dropout, precapillary arteriolar narrowing and occlusion, thereby reducing downstream perfusion[7-11]. The ensuing ischemic condition stimulates the retina and retinal pigment epithelium (RPE) to secrete substantial amounts of vascular endothelial growth factor (VEGF), leading to the disruption of the blood-retinal barrier and the proliferation of neovascularization, exacerbating the disease and resulting in irreversible worsening vision[12]. Early detection of DMI and effective management are crucial for slowing disease progression and preserving vision[5,6,13]. However, given that DMI predominantly occur within the capillaries, the detection of DMI demands the utilization of microangiography imaging modalities including fluorescein fundus angiography (FFA) and optical coherence tomography angiography (OCTA).

FFA, a time-honored technique for evaluating the retinal vascular system, is often considered the golden standard for diagnosing DMI[14-16]. This method involves the intravenous administration of fluorescein dye followed by a sequence of fundus photographs that reveal the structure and blood flow of the retinal vessels, identifying microaneurysms, areas of non-perfusion, and neovascularization[17-20]. Despite its diagnostic accuracy, FFA is an invasive procedure that carries the risk of allergic reactions, which limits its use for early screening[21,22]. OCTA is an emerging, non-invasive imaging technology that can capture the retinal capillary network and provide quantitative assessment[23-25]. It is highly sensitive in detecting DMI-related microvascular changes such as the enlargement of the FAZ, reduction in capillary density, and presence of non-perfusion zones[26-28]. However, OCTA requires expensive equipment that is not widely available, making it less feasible for screening DMI[23,29].

CFPs, in contrast to the aforementioned imaging modalities, are more ubiquitous and accessible in healthcare facilities at various settings, with a relatively simple operational procedure[30,31]. As a non-invasive examination, CFPs inflict no direct ocular trauma on patients and avoid the use of contrast agents that could elicit adverse reactions. Besides, the lower cost of CFPs translates to more affordable assessments, making it a more viable option for screening ocular diseases[32-35]. However, in clinical practice, CFPs are not currently regarded as a viable detection tool for DMI. This is likely because CFPs are thought to have limitations in rationales, including the insufficient resolution to subtle retinal microvasculature changes.

Multispectral imaging (MSI), in contrast to the color imaging technique employed in CFPs, provides superior spectral resolution. While primary color imaging captures images in three colors (red, green, and blue), MSI operates by collecting data across more wavelengths within a specific spectral range. In other words, it can distinguish between finer details in the spectral characteristics. Pigment molecules such as hemoglobin, melanin, and macular pigments present in fundus have unique spectral signatures. MSI is capable of detecting slight variations in their reflection patterns as a result of changes in the tissue's physiological state[36-39]. Having said that, MSI has not been used for studying DMI. The main hurdles could be the high cost and limited availability of fundus MSI devices. Previously, to address the accessibility issue of MSI, we developed computational multispectral imaging (CMI) methods to reconstruct multispectral data from color images, without the need for additional hardware or modifications, which provides a possible solution to reconstruct multispectral fundus images from CFPs[40-42].

After CMI, DMI detection from reconstructed multispectral fundus images entails further image processing steps. Considering the rich diversity of pigment molecules within the fundus tissue, this process inherently constitutes a multivariate analysis task. In our prior study, we had adopted the

Beer-Lambert Law to delineate tissue-optical interactions and thereby formulated multivariate equations for a semi-qualitative analysis[40,42,43]. Nevertheless, when applied to tasks demanding a higher level of precision, such as disease detection, this methodology encounters some significant challenges. Firstly, the previous method was founded upon simplified tissue models which often prove insufficient in fully accounting for the complexity of variables present in real-world human tissue. Besides, errors that occur during CMI can be amplified within the simplified linear multivariate equation system, negatively impacting the detection performance. Hence, there is an urgent need to develop a new data processing method that can robustly exploit the reconstructed multispectral fundus images to generate precise predictions for disease detection.

In this study, we developed the GNN-MSVL model with a jumper-connection mechanism to detect DMI from CFPs, adhering to several fundamental principles (**Fig.1**). First, leveraging the hypothesis that MSI can more accurately capture microvascular abnormalities in DMI, we employed CMI to reconstruct multispectral fundus images from CFPs and utilized ResNeXt101 as an encoder to extract features from multispectral views. Second, drawing on the rationale of multispectral joint analysis, we designed a GNN to analyze features from multispectral views. Additionally, we designed a jumper-connection strategy to create a cross-spectral graph structure to balance learning efficiency and comprehensiveness. In a dataset comprising 1,078 eyes of 592 patients with DM, including 530 DMI-affected eyes, our proposed method demonstrated a significant enhancement compared to the baseline model and human experts, making itself a useful tool for identifying DMI in an accessible and cost-efficient approach.

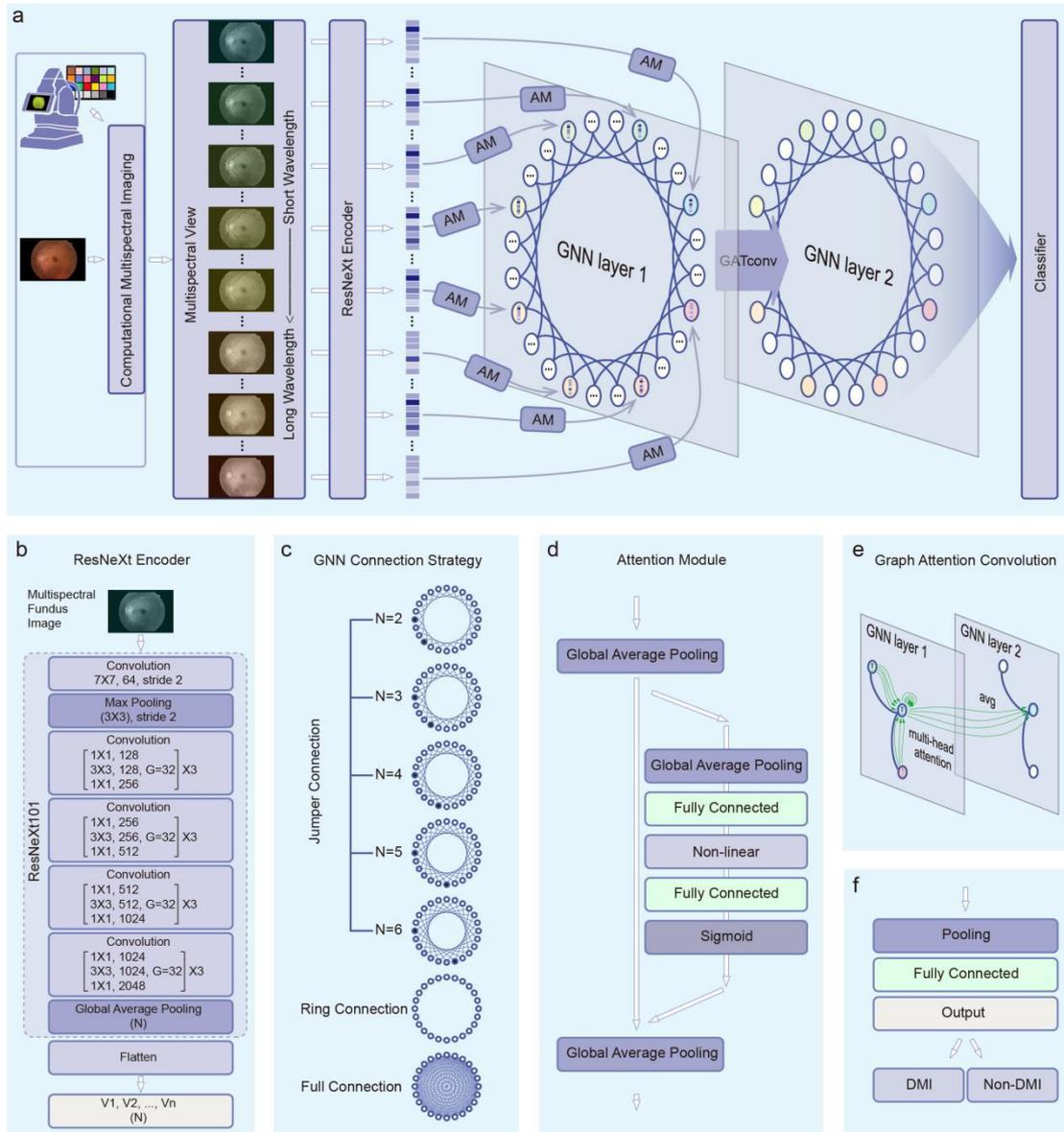

**Fig.1 Overview of GNN-MSVL model. a**. The overview of model. In CMI module, a camera calibration step is conducted to produce a transformation matrix to transform CFPs into multispectral fundus images, which are processed as multiple views of fundus in spectral space. A ResNeXt101 encoder module is used to extract representation features of multispectral views. In each spectral channel, the features are processed with an attention module (AM) to be the nodes of GNN layer 1. Among the total of 24 nodes, a jumper connection is applied to connect with each other. The learned representations of all nodes are then processed with a graph attention convolution module (GATconv) and passed to GNN layer 2. The node features of GNN layer 2 were then passed to the classifier for final prediction of "DMI" or "Non-DMI". **b**. Details of the ResNeXt101 encoder module. The encoder module takes multispectral images as inputs and processes them through convolutional layers with different configurations, including initial large-kernel convolutions and subsequent ones with varying kernel sizes and channel numbers. It features residual connections to address the vanishing gradient issue, and its output is a set of high-level feature representations that are passed to subsequent GNN layers for further processing. **c**. Illustration of the jumper connection strategy. The ring and full connection are selected as the references of minimal and maximal connection. In jumper-connection, we tested N=2,3,...,6, which means connecting every N nodes in the ring until all the nodes have been connected. **d**. Details of the attention module (AM). In each view channel, AM receives features extracted by the encoder from multispectral fundus images. The structure consists of two global average pooling layers, two fully connected layers, a non-linear layer, a sigmoid activation layer. AM weights the input features to enhance the effectiveness of feature utilization before they are passed to the subsequent GNN layers. **e**. Details of the graph attention convolution (GATconv) module. GATconv serves as a connection between GNN layer 1 and GNN layer 2, using a four-head attention mechanism for feature aggregation. **f**. Details of the classifier module.

## Results

*Computational multispectral fundus imaging*

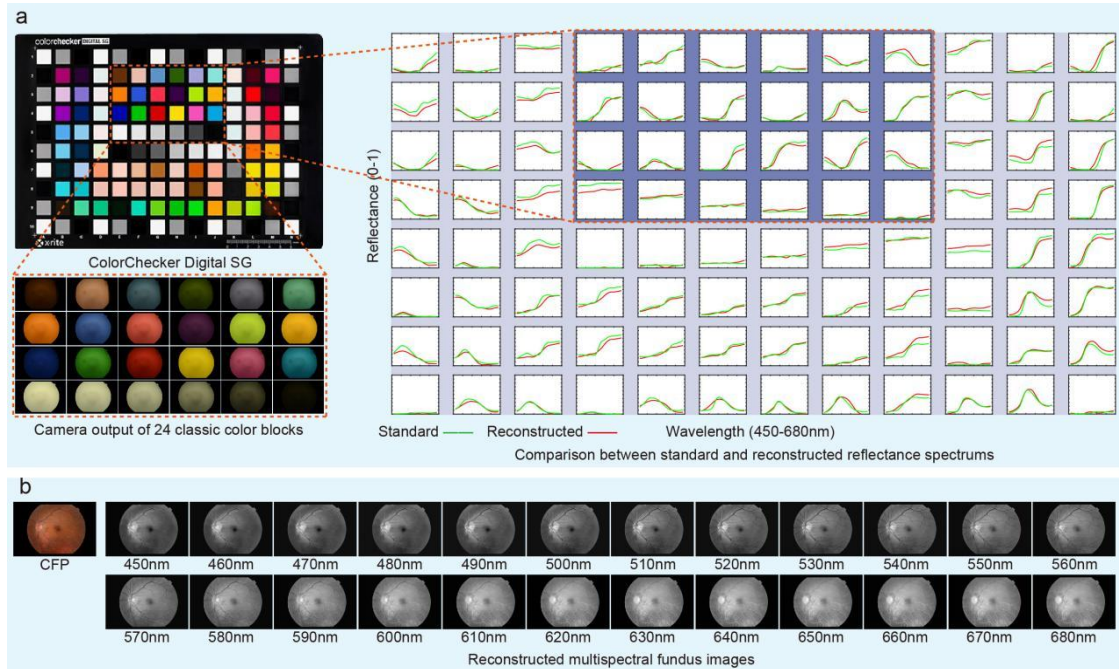

**Fig.2 Multispectral reconstruction performance of CMI. a**. Fundus camera calibration and comparison between standard and reconstructed reflectance spectra of color blocks on the ColorChecker Digital SG. **b**. Multispectral fundus images at 24 wavelengths from 450 to 680 nm reconstructed from CFPs using CMI.

**Fig. 2a** shows the camera calibration step for CMI. We captured images of the 24 classic color blocks on the ColorChecker Digital SG from the color fundus camera used in the retrospective data set collection. The resultant transformation matrix was calculated from the camera responses and then validated against a comprehensive set of 96 color patches. Comparative analysis of the standard and reconstructed reflectance spectra demonstrated CMI technique's capability to accurately reconstruct the reflectance spectra for both the reference color patches (enclosed within a blue box) and the remaining color patches on the chart. The average root mean square error (RMSE) between the standard and reconstructed reflectance spectra was calculated to be 0.05, which means the reconstruction error stays at 5% to the whole reflectance strength. A detailed breakdown of the values for each color patch is provided in **Supplementary Table s1**.

Employing the derived transformation matrix, we transformed CFPs into multispectral fundus images, spanning 24 discrete wavelengths ranging from 450 to 680 nm with a spectral resolution of 10nm (**Fig. 2b**). Our observations indicated that as the wavelength increased from shorter to longer, there was a corresponding increase in fundus reflectance, attributed to the reduction in tissue absorbance and scattering. Furthermore, the choroidal blood vessels became increasingly discernible in the images at longer wavelengths, a phenomenon attributed to the enhanced penetration depth of light at these wavelengths.

*Model performance in DMI detection*

**Table 1.** Clinical characteristics of included participants

| Characteristic | Training | Validation | Test | With DMI | Without DMI |
|---|---|---|---|---|---|
| Image level | 808 | 54 | 216 | 530 | 548 |
| Eye level | 808 | 54 | 216 | 530 | 548 |
| Participant level | 451 | 29 | 112 | 300 | 292 |
| Age, y (mean, SD) | 70.5, 11.7 | 60.9, 13.7 | 56.1, 13.4 | 67.6, 11.5 | 67.1, 15.1 |
| Male No. (%) | 52.8 | 69.0 | 56.3 | 63.1 | 45.5 |
| DMI eye No. (%) | 50.8 | 51.7 | 50.0 | | |

| DR Grading* | 0 | 1 | 2 | 3 | 4 |
|---|---|---|---|---|---|
| Eye No. in total | 507 | 81 | 412 | 37 | 41 |
| Eye No. in training data | 356 | 66 | 341 | 21 | 24 |
| Eye No. in validation data | 23 | 5 | 17 | 4 | 5 |
| Eye No. in test data | 128 | 10 | 54 | 12 | 12 |

*0: No apparent retinopathy; 1: Mild nonproliferative DR; 2: Moderate nonproliferative DR; 3: Severe nonproliferative DR; 4: Proliferative DR.

Table 2. The eye-level performance of deep learning models in the testing dataset

| | Accuracy | Sensitivity | Specificity | F1Score | AUROC | AUROC-95%CI | Cut-off | P value |
|---|---|---|---|---|---|---|---|---|
| Ophthalmologist 1 | 75.0% | 49.1% | 100.0% | 0.658 | | | | |
| Ophthalmologist 2 | 71.3% | 41.5% | 100.0% | 0.587 | | | | |
| Ophthalmologist 3 | 73.1% | 45.3% | 100.0% | 0.623 | | | | |
| CFP | 61.6% | 66.0% | 57.3% | 0.628 | 0.632 | 0.564 - 0.696 | 0.515 | Ref |
| CMI min (660nm) | 57.9% | 96.2% | 20.9% | 0.692 | 0.635 | 0.567 - 0.699 | 0.495 | P = 0.95 |
| CMI mean | 72.4% | 67.2% | 77.4% | 0.704 | 0.758 | 0.697 - 0.812 | 0.465 | P < 0.01 |
| CMI max (560nm) | 82.9% | 71.7% | 93.6% | 0.804 | 0.867 | 0.815 - 0.910 | 0.465 | P < 0.01 |
| GNN ring | 82.4% | 80.2% | 84.5% | 0.817 | 0.859 | 0.805 - 0.902 | 0.455 | P < 0.01 |
| GNN full | 82.4% | 77.4% | 87.3% | 0.812 | 0.880 | 0.829 - 0.920 | 0.495 | P < 0.01 |
| **GNN jumper (N=2)** | **84.7%** | 88.7% | 80.9% | **0.851** | **0.900** | **0.852 - 0.937** | 0.374 | P < 0.01 |
| GNN jumper (N=3) | 84.3% | 79.2% | 89.1% | 0.832 | 0.861 | 0.808 - 0.904 | 0.354 | P < 0.01 |
| GNN jumper (N=4) | 82.4% | 70.8% | 93.6% | 0.798 | 0.854 | 0.800 - 0.898 | 0.485 | P < 0.01 |
| GNN jumper (N=5) | 84.3% | 77.4% | 90.9% | 0.828 | 0.897 | 0.848 - 0.934 | 0.566 | P < 0.01 |
| GNN jumper (N=6) | 83.3% | 75.5% | 90.9% | 0.816 | 0.882 | 0.831 - 0.921 | 0.525 | P < 0.01 |

**Table 1** details the characteristics of the training, validation, and test datasets, while **Table 2** shows the performance metrics of AI models and ophthalmologists. In the test dataset, the baseline ResNeXt101 model trained on CFPs had an accuracy of 61.6%, sensitivity and specificity of 66.0% and 57.3% respectively, and an AUROC of 0.632. The ResNeXt101 models trained on multispectral fundus images exhibited average performance metrics of 72.4% accuracy, 67.2% sensitivity, 77.4% specificity, and an AUROC of 0.758. Among all wavelengths, the model trained from 560nm images achieved the best performance, with an accuracy of 82.9%, sensitivity of 71.7%, specificity of 93.6%, and an AUROC of 0.867. The ROC curves of dections made by all AI models were presented in **Fig.3**. **Supplementary Table S2** provides a detailed analysis of the performance indices for models trained on images at various wavelengths. Generally, models trained from images in the 500-600nm wavelength range performed better, likely due to the relatively higher hemoglobin absorbance in this range [**Supplementary Figure S1**].

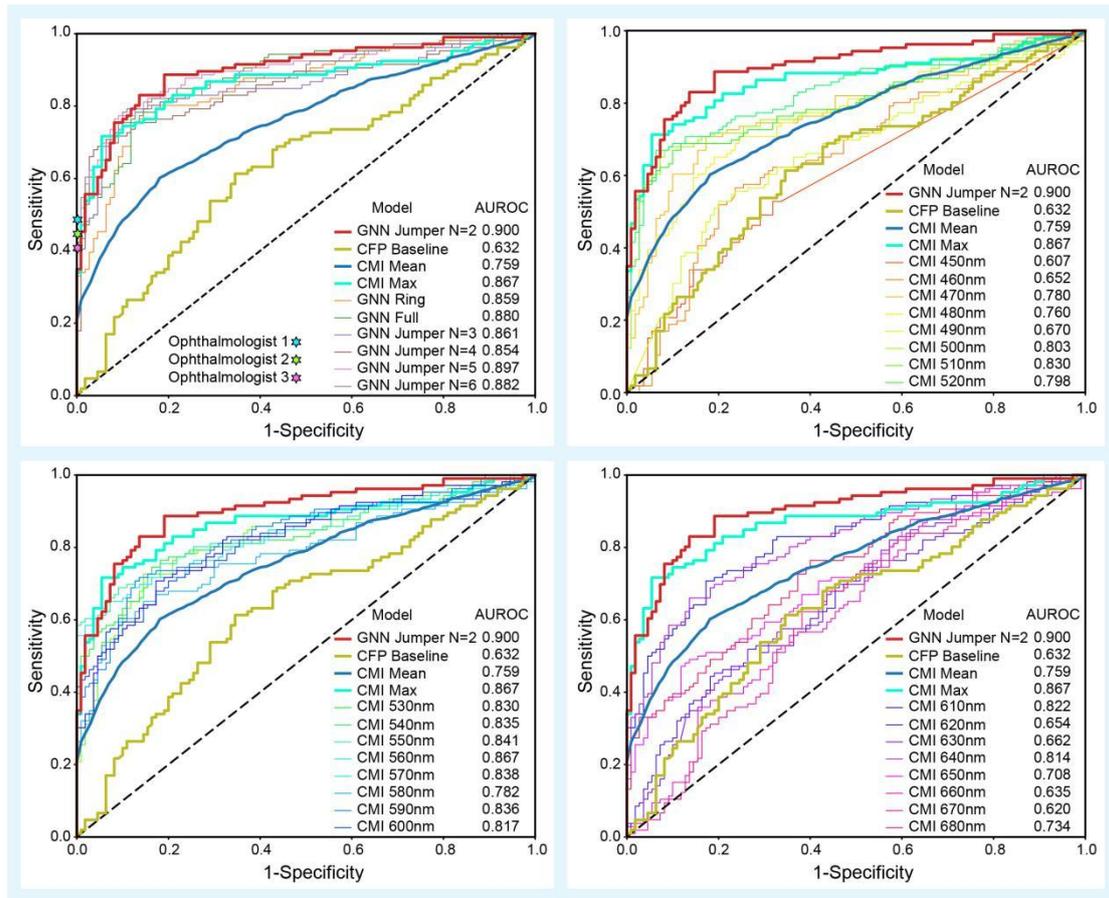

**Fig.3 ROC curves of detection made by different models.** CFP Baseline: The ROC curve of ResNeXt model trained with CFP. CMI Mean: The averaged ROC curve of ResNeXt models trained with 24 multispectral fundus images. CMI Max: the ROC curve of ResNeXt model with the highest area under the receiver operating characteristic curve (AUROC). GNN Jumper N=n: The ROC curve of GNN model with jumper connection and step width at n. CMI n nm: The ROC curve of ResNeXt model trained with multispectral fundus images at n nm.

We also evaluated GNN-MSVL models with different connection designs: full connection, ring connection, and jumper connection with step widths from 2 to 6. The GNN-MSVL model with ring connection had performance metrics of 82.4% accuracy, 80.2% sensitivity, 84.5% specificity, and an AUROC of 0.859. The full connection GNN-MSVL model achieved similar performance metrics: 82.4% accuracy, 77.4% sensitivity, 87.3% specificity, and an AUROC of 0.880. In contrast, GNN-MSVL models with jumper connections showed enhanced performance, with AUROC ranging from 0.854 to 0.900. Notably, the GNN-MSVL model with a jumper connection and a step width of 2 had the optimal performance, reaching 84.7% accuracy, 88.7% sensitivity, 80.9% specificity, and an AUROC of 0.900.

Besides the predictions of deep-learning models, a human grading experiment was conducted to comprehensively evaluate and compare different diagnostic metrics among ophthalmologists and developed AI models. Three ophthalmologists, possessing 4, 6, and 10 years of experience respectively, achieved accuracy rates of 75.0%, 71.3%, and 73.1%, sensitivities of 49.1%, 41.5%, and 45.3%, while maintaining a specificity of 100%. It's worth noting that in standard clinical settings, DMI is often believed to be undetectable from CFPs, which may account for the relatively low sensitivity in this study. Overall, this experiment provides a significant reference, emphasizing the improved performance of our GNN-MSVL model.

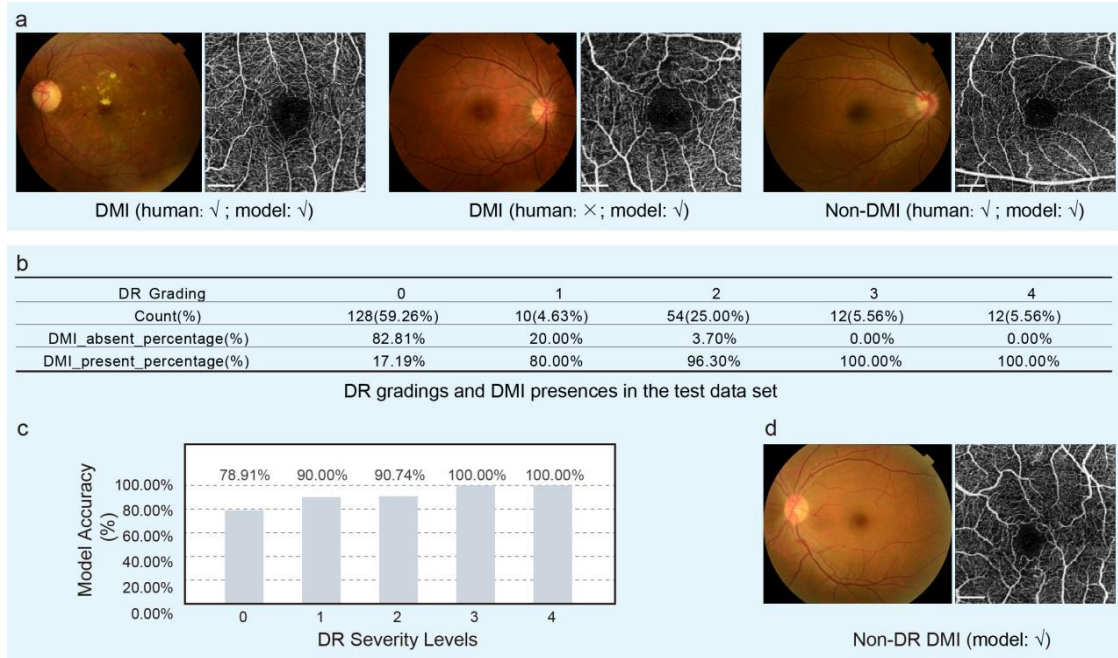

**Fig.4 Case and sub-group analysis. a.** CFP and OCTA images images of representative eyes. Predictions were made by a human grader (ophthalmologist 3 with 10 years of experience) and the GNN-MSVL model (with jumper N = 2). Left: A DMI-positive eye for which both the human grader and the model correctly predicted the outcome; Middle: A DMI-positive eye where the human grader made an incorrect prediction, while the model's prediction was correct; Right: A DMI-negative eye for which both the human grader and the model correctly predicted the outcome. **b.** DR grading and DMI presences in the test data set. **c.** Accuracy of predictions made by GNN-MSVL model (jumper N=2) across diverse DR severity levels. **d.** CFP and OCTA images of an example DMI eye without DR while GNN-MSVL model(jumper N=2) made the right prediction. Scale bars: 0.5mm.

To visually present the ability of GNN-MSVL model to detect subtle variations that human experts might miss, we selected several example eyes and showed them in Fig. 4. **Fig. 4a** displays representative CFP and OCTA images where both the human grader and the model accurately predicted DMI-positive (left) and DMI-negative (right) eyes, and sample images of DMI-positive (middle) eye with prediction discrepancies, in which the model was correct. In the OCTA images, DMI-affected eyes exhibit a wider FAZ and sparser capillaries. In contrast, from the CFP images, it is difficult to observe significant and universal signs related to the biomarkers, regardless of whether the human grader's prediction was correct or incorrect.

Since DMI is more likely to occur in patients with severe DR, we needed to rule out the possibility that the model was merely learning severe DR features instead of DMI-specific ones. To address this, we conducted a sub-group analysis based on DR grading (**Fig. 4b**). As the DR grading increases from 0 to 4, the severity of DR escalates, and the percentage of DMI presence also rises. However, the GNN-MSVL model demonstrated high detection accuracy across all five sub-groups (**Fig. 4c**). This indicates that the model is indeed learning DMI biomarkers rather than just severe DR characteristics. In **Fig. 4d**, we presented CFP and OCTA images of a DMI-positive eye without DR. The GNN-MSVL model correctly predicted the presence of DMI in this case, despite the absence of any associated DR symptoms, visually corroborating our earlier findings.

**Discussion**

In this study, we developed a novel GNN-MSVL model to accurately detect DMI from CFPs. To the best of our knowledge, this represents the first deep-learning model capable of achieving this specific task solely using CFPs . Our model integrates CMI to augment the spectral resolution of CFPs, enabling the capture of minute variations in reflectance induced by ischemic fundus microvascular tissue. By employing ResNeXt101 as the backbone for multi-view learning, it effectively extracts discriminative features from the reconstructed multispectral fundus images. Moreover, the designed GNN with a jumper connection strategy optimizes cross-spectral relationships, facilitating comprehensive and efficient multispectral-view learning. We speculate that jumper connection is superior to other connections as it reduces overfitting by having fewer connections compared to full connection and offers more flexible information integration than ring connection. A jumper step of 2 leads to the best performance may because it captures local spectral details crucial for detecting fine grained changes in DMI detection without being overly complex or missing key information[44].

CFPs have emerged as preferred tools for screening ocular diseases due to the cost-effectiveness, widespread availability, and ease of operation. For instance, it can clearly depict characteristics associated with vision-threatening diseases like DR[45,46], glaucoma[47,48], age-related macular degeneration[49,50] and papilledema[51-53]. Recent studies have demonstrated that the integration of CFPs with deep-learning can be applied in detecting a variety of eye diseases[54-57]. Particularly, CFPs have been extensively researched and applied in the screening of DR among patients with DM. However, DMI, another retinal microvascular complication caused by DM, has been historically believed to be challenging to detect from CFPs. The major reason is that the ischemic signals in DMI start from the capillary network, resulting in relatively weak and early manifestations that are not readily observable in CFPs. Our proposed model focuses on this unmet gap by enhancing the capability to assess the condition of pigment molecules, particularly hemoglobin in this study, through multispectral reconstruction and analysis. Simultaneously, through the utilization of multi-view learning and GNN, the model robustly and precisely correlates relevant information present in multispectral fundus images with DMI predictions. This strategy not only provides a potential solution for early screening of DMI but also offers an explorative path for detecting other diseases and symptoms related to subtle fundus anomalies through CFPs.

In terms of method design, our research benefits from a novel strategy termed knowledge-informed AI[58]. This paradigm differs from most of the prevailing data-driven models, which often lack a deep correlation between model design and the intrinsic semantics of the data, thereby impeding targeted optimization of deep learning models. Knowledge-informed AI leverages prior knowledge, encompassing artificial comprehension of data, the principles governing the physical world, and expert experience, to craft the model structure with certain purposes[59,60]. In our study, the conception and integration of CMI, MSVL, and GNN with jumper connections are all executed in accordance with the interaction principle of tissue optics and pathological knowledge. This informed development strategy has yielded a model that outperforms traditional data-driven models significantly, underscoring the potential of knowledge-informed AI in specific applications.

Our study has some limitations. Initially, during the CMI phase, the absence of clinical fundus MSI equipment necessitated that our spectral reconstruction verification be conducted on non-human tissue samples. Despite the inclusion of 14 representative pseudo-skin tissue samples in our dataset, potential structural disparities between these samples and actual fundus tissue may cause domain shift issues in the multispectral reconstruction, potentially introducing errors in the multispectral fundus image reconstruction process. Secondly, we opted for a spectral reconstruction band ranging from 450 to 680 nm, segmented into 24 channels with a spectral resolution of 10 nm. The selection of this spectral range was primarily aligned with the sensitivity bands of RGB cameras, while the determination of spectral resolution and channel number was somewhat arbitrary. Owing to the computational constraints associated with training and testing, we did not fine-tune the spectral resolution and the number of spectral channels, which may have precluded us from identifying the optimal spectral combination for achieving superior detection performance. Lastly, while our GNN-MSVL model with jumper connections demonstrated significantly enhanced detection performance compared to the baseline model, there remains room for further optimization of the connection strategy, attention mechanism, and classifier module. It is plausible that a more comprehensive optimization process could reveal a model structure that offers an even more accurate detection. These considerations highlight areas for future research and potential refinements to enhance the model's efficacy in detecting DMI from CFPs. In light of the aforementioned limitations, our future work will be directed towards several key areas of improvement. Initially, we aim to optimize the spectral resolution, the number of channels in the multispectral views, the connectivity patterns, the attention mechanisms, and the classifiers module, to achieve superior detection performance for DMI. Additionally, we plan to leverage the full application potential of our method by extending our research to encompass the detection of a broader spectrum of eye diseases that involve vascular and tissue pigment anomalies. Furthermore, we will explore more innovative strategies, such as developing new knowledge representations at feature engineering stages and adding physical residuals into loss functions, for integrating a more comprehensive array of prior knowledge and information into the development of deep learning algorithms. This will not only enhance the capabilities of our models but also contribute to the evolution of knowledge-informed AI methodologies within the field of ophthalmology.

In conclusion, this study proposes an GNN-MSVL model for DMI detection from CFPs. By integrating CMI to enhance spectral resolution, devising MSVL for multispectral joint analysis, and creating a ring GNN architecture with jumper connections, the model achieves high-accuracy performance in DMI detection. This achievement validates the viability and reliability of AI-based CFPs analysis in DMI detection, overcoming the long-standing skepticism in the field regarding the use of CFPs for this purpose. The successful application of this model will not only offer a more accessible and cost-effective method for DMI screening but also have far-reaching implications for early detection and intervention of eye diseases, potentially improving patient outcomes.

**Methods**

*Study design*

In this study, a deep learning model was developed for DMI detection. The macula-centered CFPs were retrospectively collected from a well-defined cohort consisting of 530 eyes diagnosed with DMI and 548 disease-free eyes. All the participants were subjected to OCTA via a swept-source optical coherence tomography (DRI OCT Triton; Topcon). A 3×3-mm volumetric scan centered at the fovea was carried out. The ground truth was labeled on the OCTA images solely by referring to the Early Treatment Diabetic Retinopathy Study protocol which is based on dye-based angiography. Specifically, the presence of DMI was identified when the images showed a disrupted FAZ, either with or without additional regions of capillary nonperfusion in the macula. In contrast, the absence of DMI was defined as images presenting an intact FAZ outline and a normal distribution of vasculature.

The CFPs were performed using a fundus camera (TRC 50DX; Topcon) which was calibrated following standardized procedures. Each CFP had a 45-degree field of view. The patients in the cohort were characterized in terms of age, gender, and relevant clinical variables[61,62]. The study was meticulously executed at the CUHK Eye Centre and adhered to strict ethical guidelines. It received approval from the Joint Chinese University of Hong Kong-New Territories East Cluster Clinical Research Ethics Committee in the Hong Kong Special Administrative Region of China, as well as the local research ethics committees of each participating center. Informed consent was obtained from each participant or their legal guardian. The consent process involved providing detailed information and obtaining signatures or electronic confirmations. The STARD guideline was scrupulously implemented for the reporting of this study, ensuring the highest level of methodological rigor, transparency, and reproducibility in the presented research findings.

*Model development*

The proposed deep-learning model consists of three notable innovative parts: (1) CMI to reconstruct multispectral fundus images from CFP, (2) MSVL to jointly analyze multispectral images, and (3) GNN with a jumper connection to achieve an efficient MSVL.

CMI: The detailed methodology of CMI is elaborated in our previous study[40,42]. Herein, we briefly outline the operation steps for applying CMI in CFP. First, we selected a classic color chart (X-rite ColorChecker Classic, X-rite Inc) as the standard color chart to calibrate the fundus camera. Specifically, we captured images of 24 color blocks present on the color chart under darkroom conditions, illuminated exclusively by the built-in flash of the fundus camera. Following the acquisition of the camera responses, we deployed the Wiener estimation algorithm to compute the transformation matrix, which facilitates the conversion of RGB values into a 24-channel reflectance spectrum. This matrix was rigorously tested against all 24 classic color blocks and the 72 non-classic color blocks on another color chart (X-rite ColorChecker Digital SG, X-rite Inc) to ascertain its accuracy. The validated transformation matrix was subsequently applied to reconstruct macula-centered CFP images into multispectral images and utilized for further analysis.

MSVL: CMI facilitates high-spectral-resolution representations of the fundus tissue, which is determined by the content and distribution of various pigments, including hemoglobin, melanin, macular pigment, and beyond[63]. Theoretically, DMI presents as ischemia in the macular region, characterized by an abnormal deficiency in hemoglobin content. Viewing tissue pigments as the independent variable set, the detection of alterations in specific independent variables necessitates the presence of a corresponding dependent variable set to accomplish multivariate joint analysis. In our research, multispectral fundus images provide the dependent variable set, serving as the foundation for joint analysis. Multi-view learning is conventionally employed to select distinct features, modalities, or perspectives of data, forming various views that reflect partial characteristics of the object, which are then integrated to obtain a comprehensive representation[64]. In this case, multispectral fundus images can be considered as views of the fundus within the spectra space, capturing the optical characteristics of fundus tissue across varying wavelengths[65]. Based on this rationale, we chose to design an MSVL framework to jointly process the spectral domain information to detect DMI.

GNN-MSVL: There are multiple ways of view fusion in multi-view learning. Among them, GNN can naturally construct multi-view data into a graph structure, regarding each view as a node and defining the relationships between views as edges, thus accurately capturing the complex topological structures and node associations[66]. Therefore, we employed GNN as the basic structure for view fusion to jointly analyze multispectral images. Furthermore, referring to the solution process of regular multivariate analysis, we also require sufficient differences among multispectral data to avoid multicollinearity.

Based on the above requirements, we designed a dual-layer GNN with jumper connections for multispectral view fusion and analysis. Specifically, from multispectral views, we used ResNeXt101 as the backbone encoder to extract view features, which, after passing through the attention module, form the nodes of GNN layer 1. To ensure spectral differences, jumper connections are adopted between nodes, that is, there are adjustable step lengths between two connected nodes. During the learning process, nodes aggregate multivariate information from neighboring nodes (different view data) and update their own representations, so that the fused features possess both the characteristics of each view and collaborative information. A GATconv module with a four-head attention mechanism aggregates all fused features in layer 1 and passed to layer 2. After processing the representations of all nodes, a binary classifier made a prediction on whether the input belongs to DMI eyes or not.

*Statistical analysis*

RMSE was used to evaluate the errors between the reconstructed spectra and the reference spectra. To assess the performance of different models for DMI detection, the precision, sensitivity, specificity, F1 score, and AUROC were calculated and compared. Additionally, we calculated the P-value of AUROC to evaluate the significance level. A P-value less than 0.01 was considered to indicate a highly statistically significant difference. We calculated Yoden's index to find the optimal cutoff point to classify test data into binary categories.

*Data and code availability*

The data used for model development of this study are not publicly available by hospital regulations to protect patient privacy. Limited data access is obtainable upon reasonable request by contacting the corresponding author. The framework is implemented in PyTorch framework 2.0.0 on an NVIDIA A100 GPU. Code and model weights are obtainable upon reasonable request by contacting the corresponding author.